# Reducing current crowding in meander superconducting strip single-photon detectors by thickening bends

**Jia-Min Xiong**[1,2,3], **Wei-Jun Zhang**[1,2,3†], **Guang-Zhao Xu**[1,2,3], **Li-Xing You**[1,2,3*], **Xing-Yu Zhang**[1,2,3], **Lu Zhang**[1,2], **Cheng-Jun Zhang**[1,2], **Dong-Hui Fan**[1,2,3], **Yu-Ze Wang**[1,2,3], **Hao Li**[1,2,3], **and Zhen Wang**[1,2,3]

[1]State Key Lab of Functional Materials for Informatics, Shanghai Institute of Microsystem and Information Technology (SIMIT), Chinese Academy of Sciences (CAS), 865 Changning Rd., Shanghai, 200050, China.
[2]CAS Center for Excellence in Superconducting Electronics (CENSE), 865 Changning Rd., Shanghai, 200050, China.
[3]University of Chinese Academy of Sciences, 19 A Yuquan Rd, Shi-jing-shan District, Beijing 100049, China

E-mail: †zhangweijun@mail.sim.ac.cn, *lxyou@mail.sim.ac.cn

## Abstract

To facilitate high optical coupling efficiency and absorptance, the active area of a superconducting nano/microstrip single-photon detector (SNSPD/SMSPD) is often designed as a meander configuration with a high filling factor (e.g., ≥0.5). However, the switching current ($I_{sw}$) of SNSPD/SMSPD, at which the detector switches into the normal state, is significantly suppressed by a geometry-induced "current crowding effect", where there are sharp bends in the strip. Here we propose and experimentally verify an alternative method to reduce current crowding both in SNSPD and SMSPD by directly increasing the thickness of the bends through the deposition and lift-off of a secondary superconducting film. We measure and compare the performance of SNSPDs and SMSPDs with different filling factors and bend configurations, with or without thickened bends. Improvements for detectors were observed in detection efficiency, intrinsic dark count rate, and time jitter, owing to the enhanced $I_{sw}$. Our method provides a promising way of optimizing SNSPD/SMSPD detection performance.

**Keywords:** superconducting strip single-photon detector, meander, current crowding effect, nanostrip, microstrip

## 1. Introduction

Superconducting nanostrip/nanowire single-photon detectors (SNSPD) [1] exhibit outstanding performance advantages over their counterparts in terms of high system detection efficiency (SDE), low dark count rate (DCR), low time jitter (TJ), and high maximum count rate (MCR) [2-8]. SNSPD plays an important role in applications such as quantum optics [9], single-photon ranging and imaging [10], deep-space laser communication [11], and quantum information processing [12].

In general, the SDE of an SNSPD can be expressed as SDE = $\eta_{coup} \times \eta_{abs} \times \eta_{ide}$ [1-3], where $\eta_{coup}$ is the optical coupling efficiency, $\eta_{abs}$ the optical absorption efficiency, and $\eta_{ide}$ is the intrinsic detection efficiency (IDE). To optimize $\eta_{coup}$, the active area of an SNSPD is usually designed to be a meander nanostrip with sharp 180° bends, covering an active area of 15–24 μm in diameter [2, 3]. In addition, To improve $\eta_{abs}$, nanostrips generally require a high filling factor ($f \geq 0.5$) [2].

To maximize SDE, $\eta_{ide}$ should be maximized, normally employing either low gap materials [4] or a narrower strip cross-section [13]. $\eta_{ide}$ reflects the spectral sensitivity of an SNSPD and depends strongly on the ratio $I_{sw}/I_{dep}$ [14], where $I_{sw}$ is the measurable switching current (at which the detector switches into the normal state), and $I_{dep}$ the depairing current for a Cooper pair [15]. However, the meander configuration introduces an intrinsic limitation to $I_{sw}/I_{dep}$. This phenomenon is well known as a geometry-induced "current crowding effect" [16-19]. Specifically, when current flows through the bends of a meandering strip, it tends to concentrate at the inner boundary of the bends, where the potential barrier for vortex entry is weakened, and the vortex is more likely to enter this region. Vortex entry and motion causes the bend regions to switch into a resistive state, thus reducing $I_{sw}$ in comparison with a straight strip of the same cross-section. Current crowding degrades detector performance, especially for those with high filling factors [17].

Recent advances have enabled SNSPDs with near unity SDE [2-7]. However, there are increasing research activities and application demands for SNSPDs that approach an "ideal" state by combining optimized detector figure merits in one detector simultaneously [3-5, 8, 20]. This would include high SDE, low DCR, low TJ, high MCR, and low polarization sensitivity. Solving the inherent physical limitations raised by current crowding may thus help to push such a study closer to the ultimate goal. Furthermore, in 2017 a new type of single-photon detector based on a superconducting microstrip (called SMSPD) attracted increasing attention [21-25] owing to its smaller kinetic inductance, higher working current, and a lower requirement in fabrication precision than those of SNSPDs, providing potential applications in the development of ultra-large active area detectors. The working principle of SMSPD relies on a large enough current biased to a specific ratio of $I_{dep}$ [24], at which the microstrip can detect single photons. This requires that the microstrip should display high uniformity not only in the strip geometry but also in the current distribution. Recent result [25] also shows that SMSPDs require a high filling-factor design to achieve high $\eta_{abs}$. Addressing the current crowding effect in SMSPD, therefore, becomes an important issue in its high-SDE implementation.

Until now, two main methods for reducing current crowding there have been used. One is to optimize the curvature of the bends through a rounded or elliptical bend design [26], a low filling factor structure [27], or even a spiral strip configuration [28, 29]. However, a rounded or elliptical bend design leads to a limited improvement in $I_{sw}$ [4]. A low filling factor design will affect $\eta_{abs}$, thus reducing the SDE. A spiral configuration has a drawback in that it is insensitive to detection in the middle, leading to higher accuracy of coupling alignment and a waste of a detector's active area [25]. In 2021, the other approach [30] was demonstrated using ion milling to thin the straight areas of nanostrips, leading to a thick bend region. Improved SDE and decreased DCR in comparison with a standard uniform thickness SNSPD (with an identical geometry) were observed.

In this paper, we propose and experimentally verify an alternative method for reducing current crowding both in SNSPD and SMSPD by directly thickening the bends through a secondary film deposition and lift-off process. We simulate, fabricate, and compare the performances of SNSPDs and SMSPDs with two-bend configurations and varied filling factors. Improvements are observed for detectors with thickened bends in terms of SDE, intrinsic DCR, and TJ. Our method is easy to operate without introducing additional damage to the detector. It also gives us a promising perspective on performance improvements for SNSPDs or SMSPDs with high filling factors. In addition, we qualitatively analyze and discuss the influence and limitation of bend-thickening on detector performance.

## 2. Simulations

Figure 1 shows optical simulations for the nanostrips carried out with commercial software (COMSOL Multiphysics, RF module). Figure 1(a) illustrates the relationship between simulated optical absorptance and NbN film thickness with a fixed strip width of 85 nm, and different filling factors (0.3-0.8), based on a double-side cavity structure [31] (see inset of Figure 1(a)). For a specific filling factor, optimal absorptance will appear at a specific thickness. For example, when the filling factor is 0.4, the optimal film thickness is ~8 nm, whereas when the filling factor is 0.7, the optimal thickness is ~4.3 nm. Previous studies [3, 32] have found that an increase in film thickness will significantly affect $\eta_{ide}$ of nanowires, owing to a nonlinear increase in $T_c$ of ultrathin films with film thickness [33]. A high-filling-factor design, therefore, helps reduce the optimal thickness of the film. Absorptance can thus be decoupled from $\eta_{ide}$ if we reduce current crowding in the bends. In addition, a high filling-factor structure will provide a benefit in low polarization sensitivity [25].

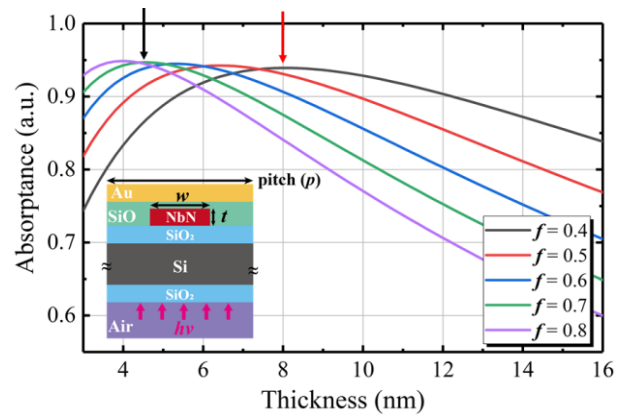

Figure 1. Optical simulations for nanostrips with a fixed width of 85 nm: absorptance vs strip thickness (*t*), with varied filling factors (*f* = *w*/*p*). Inset: cross-section schematic of an NbN strip embedded in a double-side optical cavity.

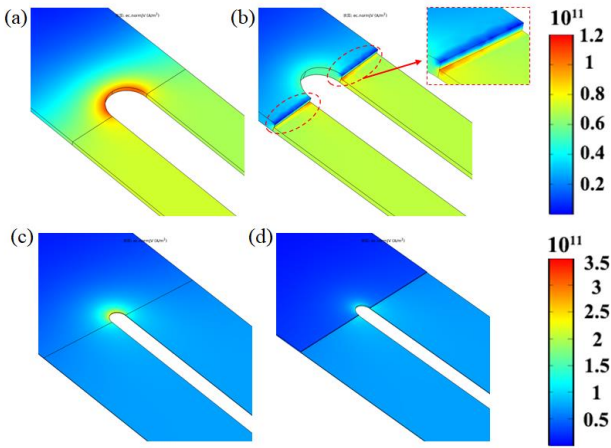

Figure 2. Numerical simulations of current crowding for a nanostrip (85 nm wide, 160 nm pitch) and microstrip (1000 nm wide, 1200 nm pitch) with a uniform thickness bend [(a) and (c)], termed "conventional", and a thickened bend [(b) and (d)], known as "bend-thickened". Different colors indicate current density magnitude. Red-dashed circles in (b) mark newly formed conners owing to a steep thickened step. Inset of (b) shows the current density distrubution of a partial enlarged view for the thickened step.

We then quantitatively evaluate current crowding in the bends of nanostrips and microstrips by simulating the current distribution for strips with different bend designs using the RF module of COMSOL Multiphysics. The results are shown in Figure 2. For comparison, we refer to a bend with a uniform thickness as a "conventional" type, as shown in Figures 2(a) and 2(c); the thickened bend is known as the "bend-thickened" type, as shown in Figures 2(b) and 2(d). The total thickness of the bend area is twice of the straight wire in this simulation. The filling factor of the nanostrip in this simulation is 0.53, with a strip width of 85 nm. Different colors in the figures indicate the relative intensity of current density ($J_c$) distributed in the strips. Current crowding is clearly significant in the conventional bend, where local $J_c$ increases in the inner edge of the curvature. In contrast, current crowding is significantly reduced in the bend-thickened structure, where $J_c$ in the inner edge of the curvature is lower than that flowing along the straight segment. Beyond the edge of the curvature, $J_c$ rapidly decreases. A thickened bend is thus no longer a bottleneck for current flowing in the meandering strip. However, we noticed that new corners form at the interface between the thickened region and straight strip segment (due to the steep thickening step, indicated with dashed circles in the figure), which shows a weak current crowding effect. Therefore, fabricating a bend-thickened structure with a smooth transition from straight segment to thick bend, will therefore help weaken current crowding at the newly formed corners.

Similar behaviors are also observed in the simulations for microstrips, as shown in Figures 2(c) and 2(d). The filling factor of the microstrip used in the simulation is ~0.83, with a strip width of ~1000 nm.

## 3. Design and Fabrication

The design and fabrication of SNSPDs and SMSPDs in this study share the same rules and fabrication processes. Take an SNSPD as an example. Figure 3 shows the layout designs of the active area for two sets of SNSPDs with two bend structures. Figure 3(a) displays the conventional bend design. Light-purple-colored areas represent regions covered with a layer of electron beam (EB) resist; darker (blue) areas represent the exposed regions, which will be exposed and etched away in a subsequent process using electron beam lithography (EBL) and reactive ion etching (RIE), respectively.

Figure 3(b) shows a bend-thickened layout, where the biggest difference to Figure 3(a) is that a green area is added, referred to as the "window region," where the secondary NbN thin film will be deposited. To ensure that the strip bend is completely covered by the thickened film (due to the limited alignment accuracy of the EBL system in the overlay process), the size of the thickened area is designed to be slightly larger than the bend area. Figure 3(c) illustrates the magnified details around the window region.

Figures 3(d)-3(l) show the main fabrication processes for our bend-thickened devices. First, in Figure 3(d), a 6.5-nm thick NbN film is deposited on a thermal-oxidized silicon

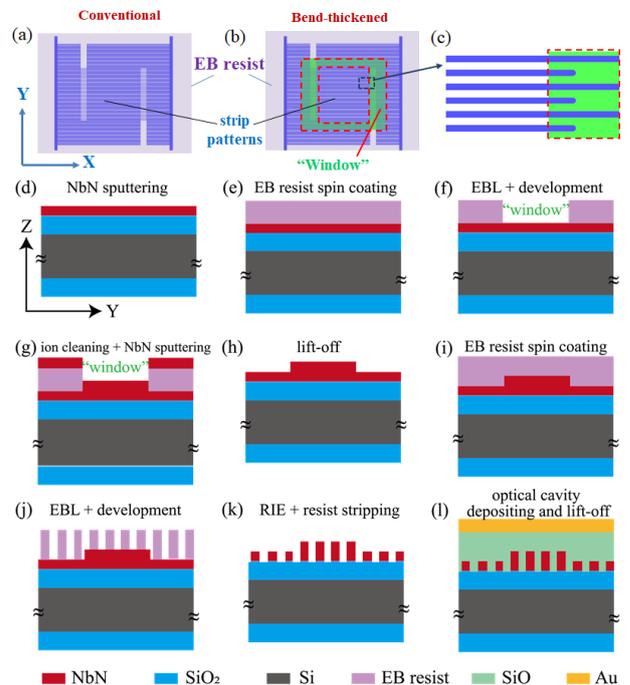

Figure 3. Design layouts of (a) standard detector, and (b) bend-thickened detector (top view). Dashed squares (filled with green) mark the "window" region. (c) Magnified picture of the bend region for (b). (d)-(l) Schematics of the main fabrication processes for a bend-thickened detector (side view).

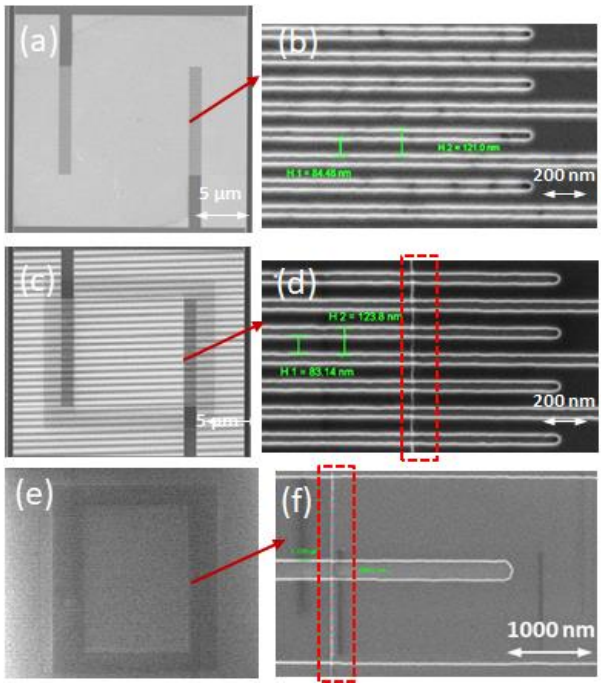

Figure 4. SEM images of fabricated SNSPDs and SMSPD with high filling factors. SNSPDs with (4a, 4b) and without (4c, 4d) bend-thickening both have a strip width of ~85 ± 5 nm and pitch ~120 nm ($f \approx 0.71$). SMSPDs with bend-thickening (4e, 4f) have a strip width of ~990 ± 5 nm and a pitch ~1200 nm ($f \approx 0.83$). Dashed boxes in 4(d), and 4(f) indicate the boundaries between the un-thickened area and the thickened area.

wafer (SiO$_2$/Si/ SiO$_2$ double-sided mirror-polished, with 268-nm-thick SiO$_2$ layers) by magnetron sputtering. A positive-tone PMMA 950A EB resist layer (~70 nm thick) is then spin-coated on the surface of the NbN film (Figure 3(e)). Through EBL and EB resist development, a window region is formed, as shown in Figure 3(f). Meanwhile, non-window regions, including the region to be fabricated with conventional devices, are all covered with the PMMA resist. The conventional and the bend-thickened devices (with the same straight segment) for comparison are fabricated under the same process on the same wafer. These two types of devices are alternately distributed on the wafer by selective exposure of the window regions.

Figure 3(g) shows a secondary NbN film disposition process, in which the oxide layer of the film in the window area is first removed by a built-in Ar$^+$ ion gun through ion cleaning (cleaning time ~60 s, under a pressure of ~6 Tor and plasma power ~100 W) in a magnetron sputtering system. An additional NbN film (9 nm or thicker) is then deposited through the magnetron sputtering process. After the secondary NbN film is prepared, the film covering the EB resist is peeled off in acetone using a lift-off process, leaving only the thickened NbN film in the window area, as shown in Fig. 3(h). Following this, the EB resist (ZEP (1:1.5) ~100 nm) is spin-coated on the wafer for further EBL. The strips are patterned by EBL and etched using RIE in a CF$_4$ plasma. An appropriate etching time (~70 s) is selected to ensure that the thickened bend area is etched through. A cross-sectional schematic of the bend-thickened structure is shown in Figure 3(k). The subsequent process is basically the same as the standard process, such as removing the residual EB resist by NMP solvent, fabricating the electrodes by ultraviolet lithography and RIE, deposition and lift-off the optical cavity (SiO layer and an Au mirror, Figure 3(l)), wafer dicing and wire bonding for measurements.

In order to test process reliability, we fabricated samples in three batches: the first and second batch of samples are nanostrip devices (NS-1#, and NS-2#), and the third batch is microstrip devices (MS-3#). Note that the nominal thicknesses of the first NbN layer for NS-1#, NS-2#, and MS-3# samples are ~6.5 nm, ~5 nm, and ~6.5 nm, respectively. In the three batch samples, the nominal thickness of the secondary NbN film deposition was set to ~9 nm. We, therefore, have dozens of devices in each batch for comparison, all fabricated on the same wafer.

Figure 4 shows typical scanning electron microscopy (SEM) images for the nanostrip (NS-2#) and microstrip (MS-3#) devices. The left panels show the entire active area of these devices, and the right panels show enlarged images of the bend area. Specifically, the first row (Figures 4(a) to 4(b)) and second (Figures 4(c) to 4(d)) correspond to conventional and bend-thickened nanostrips, respectively, where the nanostrips are fabricated with a width of ~85 nm and pitch ~120 nm; the third row (Figures 4(e) to 4(f)) corresponds to a microstrip with a bend-thickened design of width ~990 nm and pitch ~1100 nm.

In Figures 4(c) and 4(e), the thickened regions demonstrate dark-colored areas, corresponding to the "window" region shown in Figure 3(b). An obvious boundary between the first layer area and the second layer area (thickened region) are observed, corresponding to the thickened step. A step height of ~9 nm was determined using an atomic force microscope, and the step verticality was approximately 83°–87°. Through process optimization, we expect to achieve a step with a smaller inclination angle to reduce newly formed bend effects. Overall, these SEM images demonstrate that nanostrips and microstrips with high filling factors were successfully fabricated, with clear profiles in strips and bends.

## 4. Results and Analysis

### 4.1 I-V and R-T curves

We characterized the devices in a 2.2-K Gifford–McMahon (G–M) cryocooler. Figure 4 shows the typical current–voltage (IV) and resistance–temperature (RT) curves of our nanostrips (NS-1#, Figure 5(a)–(b)) and microstrips (MS-3#, Figure 5(c)–(d)) with the two-bend designs. The nanostrip devices (NS-1#A4 and NS-1#B4) are 80 nm wide and 6.5 nm thick, with an 0.5 filling factor and an active area of ~18 μm × 18 μm. The microstrip devices (MS-3#E5 and MS-3#F5) are 990 nm wide and 6.5 nm thick, with a 0.83 filling factor and an active area of ~50 μm × 50 μm. Among these two sets of IV curves, the $I_{sw}$s of the bend-thickened devices increase. The $I_{sw}$ increment is more profound in the bend-thickened microstrip device, owing to greater $I_{sw}$ suppression at a higher filling factor and wider strip width.

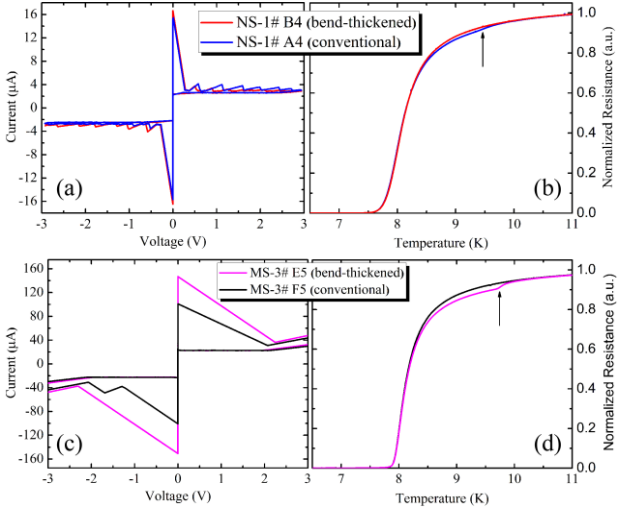

Figure 5. Current-voltage (IV) and resistance-tempeture curves for different bend types of superconducting strips: (a)-(b) nanostrip devices; (c)-(d) microstrip devices. The resistance is normalized to the normal-state resistance at 12 K.

For the R–T curves, it is interesting that a small transition step appears in the transition of the strips from a superconducting state to a normal state. These transition steps are indicated with arrows in Figures 5(b) and 5(d), where the temperature corresponding to the step inflection point is ~9.1 K. A small transition step proves that the secondary deposited NbN film merges with the initial NbN film to form a new, thicker NbN layer, since the $T_C$ of ultrathin NbN depends strongly on film thickness [33]. The transition step of the bend-thickened microstrip device is evident because the bend segments in the microstrip contribute a higher proportion to the total resistance of the entire strip compared with the nanostrip. We define the device $T_C$ as the temperature corresponding to 50% of the normal state resistance at 12 K. The $T_C$ of the 6.5-nm-thick NbN film is then ~8.14 K.

Figure 6 further shows measured $I_{sw}$ statistics for more samples. Figures 6(a) and 6(b) show the $I_{sw}$ of the samples as a function of the filling factor for nanostrip devices (NS-1#, red dots; NS-2#, black squares) and microstrip devices (MS-3#, blue triangles), respectively. For each data point, we measured at least three devices from the same wafer to ensure reliability in the results. Error bars in the figure show the standard deviation of $I_{sw}$ for the samples. We found that, in both nanostrip and microstrip devices, the $I_{sw}$ of conventional devices decreases as the filling factor increases, which is consistent with previous reports [16, 17, 19]. Bend-thickened devices show a nearly flat tendency which implies almost no dependence on changes in the filling factor, even at a very high filling factor (e.g., nanostrip NS-2# at $f$ ~0.71, or microstrip MS-3# at $f$ ~0.83). This shows that we have successfully reduced the current crowding effect in the bends of the meandering strip.

Figure 6(c) shows the experimental $I_{sw}$ enhancement factor ($f_{en}$) of our nanostrip and microstrip devices against the filling factors, respectively. $f_{en}$ is defined as the $I_{sw}$ ratio between the bend-thickened and conventional devices at the same filling factor. $f_{en}$ of the nanostrip and microstrip devices seem to follow the same trend when the filling factors change. For microstrip devices, when the filling factor is ~0.28, the current crowding shows a weak influence on $I_{sw}$ ($f_{en}$ ~1.03). This implies that meander microstrips with a low filling factor design would help to achieve a weak current suppression. A significant enhancement of the $I_{sw}$ in the bend-thickened nanostrips and microstrips with high filling factor is observed, such as $f_{en}$ ~1.5 at $f = 0.83$ for the microstrip device and $f_{en}$ ~1.4 at $f = 0.71$ for the nanostrip device. Compared with the previous works [25, 30], it can be found that the reproted $f_{en}$ as a function of $f$ follows the similar trend. However, there are still some deviation points, which may be attributed to the errors of the strip fabrication (width error, constriction, and bend curvature formed in EBL and RIE processes). We will further analyze the issues in the discussion section.

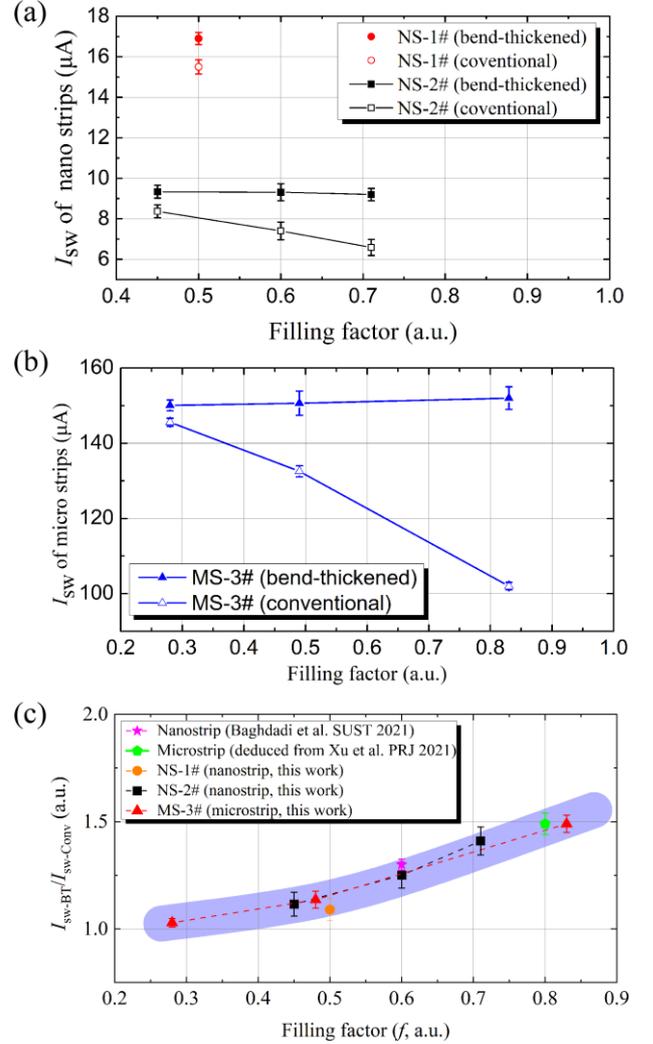

Figure 6. Summary results for the two types of strips with varied filling factors: (a) nanostrip devices (NS-1# and NS-2# with 1st layer $t ≈ 6.5$ and 5 nm, respectively; active area ~18 μm × 18 μm); (b) microstrip devices (MS-3#,1st layer $t ≈ 6.5$ nm active area ~50 μm × 50 μm); (c) enhanced ratio ($f_{en} = I_{sw-BT}/I_{sw-Conv}$) of $I_{sw}$ vs. filling factor ($f$), due to bend thickening. Symbols: star (nanostrip; $w$ ~85 nm, $f = 0.6$, active area ~5 μm × 2.6 μm, adopted from Baghdadi et al. [27]); pentagon ($f_{en}$ of the microstrip is derived from the $I_{sw}$ ratio between the spiral and meander configurations with the same active area [23]); squares (nanostrip: NS-1#, $w$ ~80 nm); circles (nanostrips: NS-2#, $w$ ~85 nm); triangles (microstrips: MS-3#, $w$ ~990 nm). The light blue area marks the possible distribution of $f_{en}$ as a function of $f$, used to guide the eyes.

*4.2 SNSPD performance*

We used a continuous-wave light source, variable attenuators, polarization controller, and a high-precise power meter to determine the SDE of the detectors [3]. A TCSPC-150 module and a femtosecond pulse light sources are used for the TJ measurement [34]. The working wavelength of the light source are all set at 1550 nm, and the working temperature of the device is ~2.2 K. Two representative detectors (NS-1#A4 and NS-1#B4) are selected for each type, with a criterion of an average $I_{sw}$ among its own type of devices.

Figure 7 shows the detector performance for the two-bend types of the devices, where SDE, intrinsic DCR, and TJ are plotted as a function of bias current. Here intrinsic DCR refers to that when the device is not coupled to an optical fiber, and light radiation is shielded. For the conventional (bend-thickened) device, the $I_{sw}$ is 15.5 (16.9) μA, corresponding to ~1.4 μA suppression due to current crowding. As shown in Figure 7(a), both sets of SDE curves show trends approaching saturation, while the bend-thickened device is slightly better. The maximum SDE (SDE$_m$) of the two detectors is ~70.3% and 73.6%, respectively. To evacuate the IDE of the device, empirical sigmoidal curves are fitted to the measured data. IDE is calculated as SDE$_m$/SDE$_{asy}$, where SDE$_{asy}$ is the asymptotic value of the sigmoid fit. In this way, we found that the IDE of conventional and bend-thickened devices was 94% and 98%, respectively. A more saturated plateau in SDE usually means that a higher temperature that a detector can be operated, and a higher SDE at the same DCR.

It is also interesting that, at the same bias current of ~15.3 μA, the SDE of the two types of devices is basically the same (~70%). However, the intrinsic DCR of the bend-thickened device is significantly reduced by nearly three orders of magnitude (i.e., from 102.7 cps to 0.1 cps as shown in the figure) compared with the conventional device. In future, if one could effectively suppress the background DCR, for example, using a cryogenic bandpass optical filter [35], the bend-thickened method would be helpful to improve the limit optical signal-to-noise ratio of the detector.

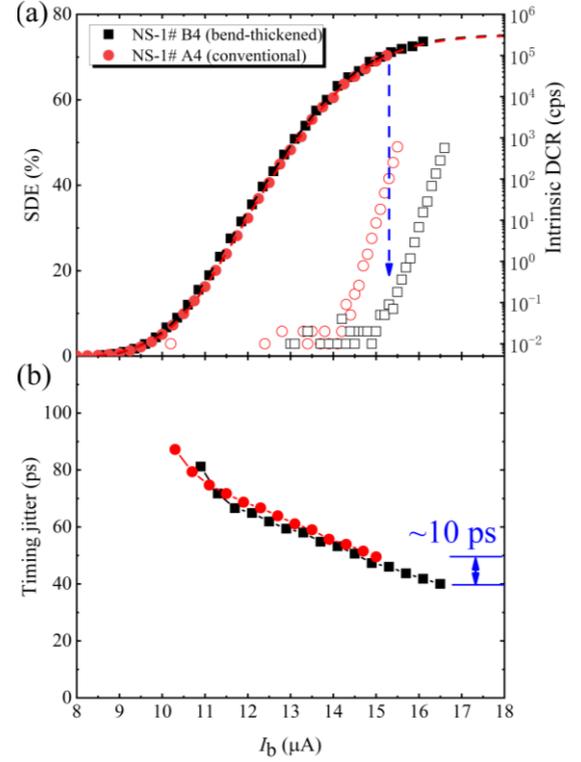

Figure 7. Detector performance of 0.5 filling-factor SNSPDs with/without thicken bends, illuminated with 1550 nm photons. (a) SDE and intrinsic DCR vs. bias current ($I_b$). (b) timing jitter vs. $I_b$. Dashed lines are empirical sigmoidal fits.

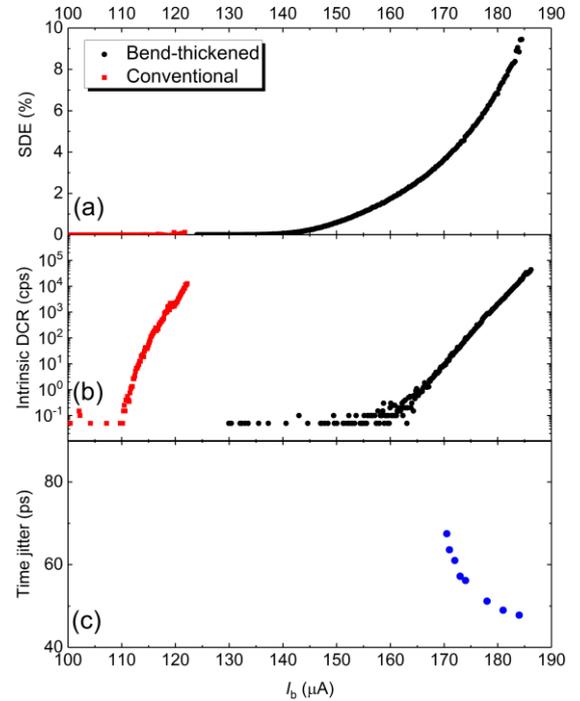

Figure 8. Detector performance of 0.83 filling factor SMSPDs with and without thickened bends, illuminated with 1550 nm photons. (a) SDE vs. $I_b$; (b) Intrinsic DCR vs. $I_b$. (c) timing jitter vs. $I_b$.

As for TJ, the two types of devices exhibit a similar current dependence, as shown in Figure 7(b), where TJ decreases as the bias current increases, owing to an improvement in the electrical signal-to-noise ratio [34]. However, owing to the $I_{sw}$ enhancement, the bend-thickened device can be biased to a larger current, thereby lessening its TJ. The timing jitter of the bend-thickened device is reduced from 50 to 40 ps.

## 4.3 SMSPD performance

We next proceed to characterize the performance of the two types of microstrip detectors (MS-3#E5 and MS-3#F5). In a different manner from the readout circuit of the nanostrip, for the SMSPDs, in order to prevent the device from latching [21, 25], we connected a 6.8 Ω shunt resistor in parallel with the device chip through wire bonding, selecting a device group with a high filling factor (~0.83) for comparison, because such a design is preferred in microstrip devices. Figure 8 shows the bias current dependence of the SDE, together with intrinsic DCR and TJ for these two sets of SMSPD structures. Surprisingly, the SDE of the conventional SMSPD is less than 0.1% due to strong $I_{sw}$ suppression by current crowding. However, the bend-thickened device demonstrates an SDE of around ~8% at 1550 nm.

The low efficiency of conventional devices prevents the measurement of its TJ. Thus, only the TJ of the bend-thickened device is measured, as shown in Figure 8(c). The timing jitter of SMSPD decreases with increased bias current, and we observe a minimum jitter of ~40 ps.

## 5. Discussions

Experimentally, we have observed that the switching current of a device with a bend-thickened design was almost independent of the filling factor. However, it is still difficult for us to say that the influence of the bend-induced current crowding effect was completely eliminated, because the switching current we currently measured could be the result of multiple factors, such as a weakening of the current crowding effect, (fabrication-induced or intrinsic) defects/constrictions [36], and measurement errors [37].

Theoretically, for the quantitative evaluation of the current crowding effect (e.g., $f_{en}$ vs. $f$), it is currently difficult to obtain a good theoretical prediction consistent with the experiment, owing to simplifications of the model [16]. Therefore, research on this aspect, especially the numerical simulation based on time-dependent Ginzburg–Landau (TDGL) equations [24], would be helpful to address this issue. In addition, based on the present experimental results, the new-formed bend effect caused by the thickened step seems to be weak. We speculate that this is because the thickened thickness (~9 nm) is comparable with the coherence length of NbN film (ξ ~ 5 nm), and the edge barrier dominates and prevents the entry of the vortices, weaking its effect on the switching current. However, this still needs further analysis via TDGL numerical simulation.

We believe that the improvement in SDE or IDE in a bend-thickened design is device-condition dependent, i.e., the improvement may be limited by fabrication accuracy, bend curvature, device material, operating wavelength, and temperature. The reasons for this are as follows. First, to some extent, the current crowding effect can be regarded simply as constrictions in the superconducting strip (but there are still some differences in the effects of bends and constrictions [38]), in which case there could be a competitive effect between current crowding constrictions and defects in the strip. When the influence of defects is dominant, the improvement due to bend-thickening would be lessened. Good device fabrication conditions would thus be necessary to make the bend-thickening functional. Second, there is bend curvature, e.g., an elliptical bend design would slightly reduce current crowding (~6% $I_{sw}$ increment) [4, 30], which may also weaken bend-thickening performance. Third, the degree of SDE or IDE improvement by bend-thickening is closely related to photon-response sensitivity (i.e., IDE at specific photon energy) of the detector under test. IDE is influenced by many factors [14], including the energy gap of strip material, strip cross-section, excitation photon energy, and operating temperature. When detectors are at IDE saturation (or close to saturation) at a specific wavelength and biased above a specific bias current, bend-thickening has a limited improving effect on SDE or IDE. For example, in Figure 6, at 1550 nm, the IDE of the two detector bend types is close to saturation, and so the IDE increase due to bend improvement is not very large (e.g., an increment of ~4% as shown in Figure 7(a)). In contrast, IDE-constrained devices (e.g., operated at longer wavelengths) or devices far away from IDE saturation (e.g., the SMSPD for 1550 nm in Figure 8, or the reported results at 500 nm in reference [30]), bend-thickening will bring significant improvements in IDE due to the direct increase in the ratio $I_b/I_{dep}$.

To further demonstrate this argument, we compare the influence of current crowding on the two most representative detectors made of amorphous MoSi and our own polycrystalline NbN material. Here we assume the enhanced factor $f_{en}$ of $I_{sw}$ shown in Figure 6(c) is universal for all strip detectors. The switching current without current crowding can thus be expressed as $I_{sw0} = I_{sw} \times f_{en}$.

Figure 9 shows the normalized current dependence of normalized DE (NDE) for two such kinds of detectors. The bias current is normalized to $I_{sw0}$, and the curves are recorded under 1550 nm photon illumination. In Figure 9, the open symbols are experimental data obtained from the literature [3, 4]. The dashed lines are empirical sigmoidal fits for the two detector types, assuming no current crowding in the strips (labeled as an ideal case). As a rough and qualitative comparison, we can see that the MoSi detector presents a wide and flat saturation plateau ($I_b/I_{sw0} > 0.6$), while the plateau of NbN is short ($I_b/I_{sw0} > 0.8$). The onset detection current of NDE (at which NDE become larger than 0.1) in the NbN detector is also higher than that of the MoSi detector. The blue arrow in the figure indicates the estimated suppressed current for different filling factors due to current crowding (i.e., $1/f_{en}$). From the perspective of NDE alone, the MoSi detector is found to be less affected by current crowding ($f = 0.9$, NDE ~0.99), while NbN is more affected ($f = 0.9$, NDE ~0.67). Therefore, the influence of NDE due to current crowding is much dependent on the photon-response sensitivity of the detector.

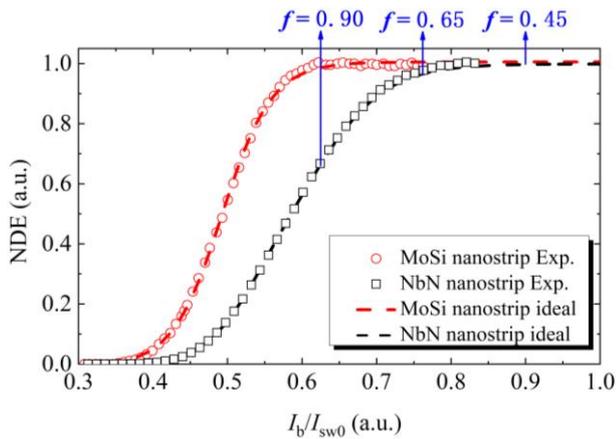

Figure 9. Typical NDE as a function of normalized current for two types of SNSPDs made of MoSi and NbN, respectively, illuminated at 1550 nm photons. Open symbols are normalized experimental data adopted from the references [3] (MoSi, 4 nm thick, 110 nm wide, and 170 nm pitch, operated at 700 mK) and [5] (NbN, 7 nm thick, 75 nm wide, and 140 nm pitch, operated at 16 mK). Dashed-lines are sigmoidal fits for the experimental data with estimated $I_{sw0}$. Blue arrows mark the possible $I_{sw}$ at different filling factors ($f$), suppressed by the current crowding effect.

Regarding DCR and TJ, improvement in the bends can bring an expected increase in $I_{sw}$, and so improvements in intrinsic DCR and time jitter are expected. For example, in Figure 9, although the increase in efficiency of the MoSi detector due to bend-thickening is not large, the enhanced ratio of the switching current is quite significant, which would help in improving its TJ and DCR. We therefore believe that a bend-thickened design can improve the performance of the detector.

Finally, we compare our method with a reported approach that employs ion milling to thin the straight strip segments [30]. Our method directly increases the thickness of the bend area through secondary film deposition to effectively reduce current crowding. Although we use ion cleaning in the bend regions (to remove the oxide layer), a thicker NbN film is formed in the secondary film deposition, which enhances the superconductivity of the bends, and this helps counteract the negative effects of ion cleaning. Our method therefore avoids the introduction of new processing defects. In the future, we will fabricate larger active area detectors with thickened bend areas to test the adaptability and scalability of this method.

## 6. Conclusions

We have systematically investigated the feasibility of reducing current crowding in meander nanostrips and microstrips through the direct thickening of bends. A significant enhancement in switching current in bend-thickened nanostrips and microstrips with a high filling factor is observed (e.g., ~1.5 times for 0.83 filling factor). We also observe that the switching current of a device with a bend-thickened design is almost independent of the filling factor. By comparing devices with or without bend-thickening treatment, we find that a bend-thickened device can improve system or IDE, intrinsic DCR, and TJ. For example, TJ is reduced by 10 ps for bend-thickened samples with $f$ ~0.5 in this study. We notice that the degree of detection efficiency improvement due to bend-thickening is condition-dependent, and this method has a more profound effect on improving the detection performance of devices with a low IDE, such as meandering microstrips with a high filling factor.

We believe that bend-thickened detectors will have potential application for high filling factor, low polarization sensitivity SNSPDs or SMSPDs with high detection efficiency and low DCR. Our method also provides an alternative way to investigate the bend-related mechanism, e.g., bend-induced TJ [29, 39], photon-response count rate [38, 40], and DCR [40].

## Acknowledgments

This work is supported by the National Natural Science Foundation of China (NSFC, Grants No. 61971409), the National Key R&D Program of China (Grants No. 2017YFA0304000) and the Science and Technology Commission of Shanghai Municipality (Grants No. 18511110202). W. J. Zhang is supported by the Youth Innovation Promotion Association (No. 2019238), Chinese Academy of Sciences.